\documentclass[traditabstract]{aa} 

\usepackage{natbib}
\bibpunct{(}{)}{;}{a}{,}{,}

\usepackage{graphicx}
\usepackage{txfonts}
\usepackage{multirow}

\begin{document}

\title{Multiwavelength flux variations induced by stellar magnetic activity: effects on planetary transits}

\author{P.~Ballerini\inst{1,2} \and G.~Micela\inst{2} \and A.~F.~Lanza\inst{3}  \and I.~Pagano\inst{3}}
\institute{Dipartimento di Fisica ed Astronomia,
          Universit\`a di Catania, Via Santa Sofia 78, 95123 Catania, Italy  \and INAF-Osservatorio Astronomico di Palermo, Piazza del Parlamento 1, 90134 Palermo, Italy \and INAF-Osservatorio Astrofisico di Catania, Via Santa Sofia 78, 95123 Catania, Italy   \\
          email:pba@oact.inaf.it}
\date{Received; accepted }
\abstract
{Stellar magnetic activity is a source of noise in the study of the transits of extrasolar planets. It induces flux variations which affect significantly the transit depth determination and the derivations of planetary and stellar parameters. Furthermore, the colour dependence of stellar activity may significantly influence the characterization of planetary atmospheres. Here we present a systematic approach to quantify the corresponding stellar flux variations as a function of wavelength bands. We consider a star with spots covering a given fraction of its disc and model the variability  in the UBVRIJHK photometric system and in the \textit{Spitzer}/IRAC wavebands for dwarf stars from G to M spectral types. We compare activity-induced flux variations in different passbands with  planetary transits and quantify how they affect the determination of the planetary radius and the analysis of the transmission spectroscopy in the study of planetary atmospheres. We suggest that the monitoring of the systems by using broad band photometry, from visible to infrared, helps  to constraining activity effects. The ratio of the relative variations of the stellar fluxes in  short wavelength optical bands (e.g., U or B) to near infrared ones (e.g., J or K) can be used to distinguish starspot brightness dips from  planetary transits in a stellar light curve. In addition to the perturbations in the measurement of the planetary radius, we find that starspots can affect  the determination of the relative semimajor axis and the inclination of the planetary orbit which have a significant impact on the derivation of the stellar density from the transit light curves.}

\titlerunning{Multiwavelength flux variations induced by stellar magnetic activity}
\authorrunning{P.~Ballerini et al.}
\maketitle

 \section{Introduction}
  \label{intro}
  The microvariability of active stars can hamper the measure of planetary transits and this can be a problem  for targets with a solar-type or slightly higher magnetic activity level, i.e. main-sequence stars from late F  to late G-type, or even worse for dM stars. The transit of a large group of spots on the solar surface causes a relative decrease of the optical flux of $\sim$3.3 $\times 10^{-3}$ \citep{Froehlich:2004} and even in the absence of sunspots, the solar flux exhibits a modulation due to bright structures, such as photospheric faculae. For comparison, the transit of a planet across the solar disc would produce a relative flux variation of the order of 10$^{-4}$ and 10$^{-2}$ for Earth and Jupiter, respectively. Therefore, the search for planetary transits in the presence of stellar activity requires techniques to filter out or fit the microvariability of the star. They are based on the observations of the flux variations outside transits along time intervals of at least two or three stellar rotations \citep[cf., e.g., ][]{Aigrain:2004,Moutou:2005,Bonomo:2008,Bonomo:2009}. 

  The detection of a transiting exoplanet is favoured at infrared wavelengths, where activity variations are smaller, especially when surveying dM-type stars. Moreover, their smaller size provides a more advantageous ratio between the planet and the host star radii, implying a greater transit depth for a given planetary radius, and suggests that for dM stars it could be easy to detect low mass planets \citep{Butler:2004,Nutzman:2008,Deming:2009b}. However a large fraction of dM stars shows a significant level of magnetic activity which may be a crucial problem for the detection of transiting planets.

  In the infrared bands, the ratio of the planet to the star emission increases remarkably, thus we can directly detect the radiation coming from the planet, allowing us to characterize the atmosphere during the occultation phase, i.e. through the analysis of the reflected planetary spectrum (emission spectroscopy, see \citealt{Charbonneau:1999,Cameron:1999}) or during the planetary transit in front of the star, i.e. by analyzing the  transmission  spectrum of the planetary atmosphere (transmission spectroscopy, see \citealt{Seager:2000,Brown:2001b,Hubbard:2001,Charbonneau:2002}). The analysis of infrared spectra is crucial for transmission spectroscopy, since the infrared domain includes the most important spectral signatures of the main atmospheric moleculae even if their detection is still controversial \citep{Tinetti:2007a,Swain:2008,Desert:2009,Sing:2009,Agol:2010,Desert:2011a,Palle:2011,Gibson:2011}.

  The observation of stars hosting planets in the optical and near-infrared wavebands is the main scientific objective of a number of ground-based projects (e.g. {\it MEarth}, see \citealt{Nutzman:2008}). It is part of the  program of the \textit{JWST} (James Webb Space Telescope, \citealt{Gardner:2006}) and will be the focus of the proposed ESA mission, \textit{EChO}\footnote{The \textit{EChO} mission has been selected for the assessment phase for the medium-class mission, to be launched in the period $2020-2022$ by ESA.} (Exoplanet Characterization Observatory, \citealt{Tessenyi:2010}), a space-borne telescope that will observe the visible-infrared spectra ($0.4-16$ $\mu$m)  to characterize the physical properties of exoplanet atmospheres, searching for molecular spectral features and bio-markers. A significant part of the mission will be devoted to the observation of the atmospheres of Super-Earth planets orbiting M dwarf stars for the aforementioned reasons.

  The most widely used approach to better constrain the planetary atmospheres is that of measuring the transit depth in different passbands observed simultaneously. The selective absorption of the stellar radiation by the molecular species, present in the planetary atmosphere, may produce the variation of the apparent radius of the planet versus the central wavelength of the passband and  can be used to detect them \citep[e.g.,][]{Desert:2011a,Desert:2011b,Croll:2011}. 
  The presence of stellar magnetic variability, due to cool spots, bright faculae, or magnetic surface inhomogeneities in general, can modify the transit depth and this can have a significant impact on the derivation of the dependence of the planetary radius on the wavelength \citep{Pont:2008,Czesla:2009,Sing:2009,Agol:2010,Berta:2011,Sing:2011a,Sing:2011b,Desert:2011a}. The issue is to understand if the apparent radius variations are due to molecular species in the planetary atmospheres or to stellar activity whose effect is also expected to depend on the bandpass. In this respect, the quantification of activity-induced variations, as functions of wavelength, may be crucial for the transmission spectroscopy and the characterization of the planet atmosphere and composition. These effects are crucial above all for active sources, such as HD~189733, but the issue rises even for less active sources, such as GJ~1214. 

  Since the discovery of its planetary companion \citep{Bouchy:2005}, HD~189733 has been the subject of intense observations both in the optical bands \citep{Bakos:2006,Winn:2007,Pont:2007,Pont:2008,Sing:2011b} and in the infrared wavebands \citep{Beaulieu:2008,Sing:2009,Desert:2011a}. It is a K-type star with a strong chromospheric activity \citep{Wright:2004} which produces a quasi-periodic optical flux variation of $\sim$1.3\% \citep{Winn:2007} due to the rotation of a spotted stellar surface. The first accurate determination of the planetary system parameters is carried out by \cite{Bakos:2006} through BVRI photometry; neglecting the effects of stellar variability, their planet-to-star radius ratio ($0.156 \pm 0.004$) results smaller than that of \citealt{Bouchy:2005} ($0.172 \pm 0.003$)\footnote{Throughout this section, we will present the results achieved in the literature by comparing the planet-to-star radius ratio of each source, so as to avoid any inconsistency due to the use of different stellar radii and to match quantities directly related to the observations. Where needed, we will compute the $R_{\rm p}/R_{\rm \star}$ ratio by using the stellar radius adopted in the relevant paper.}. \cite{Winn:2007} obtain radius measurements compatible with those of \cite{Bakos:2006}, again neglecting the effects of stellar variability. Even \cite{Pont:2007} obtain results compatible with those of \cite{Bakos:2006}, in this case correcting their \textit{HST}/ACS observations for the stellar variability; moreover these authors expect that the starspots effect be reduced in the infrared because of a lower spot contrast. Through \textit{HST}/NICMOS transit observations, correcting for the unocculted starspots, \cite{Sing:2009} find a planet-to-star radius ratio of $0.15464 \pm 0.00051$  and $0.15496 \pm 0.00028$ at 1.66 and 1.87 $\mu$m, respectively, in agreement with the planet-to-star radius ratio of \cite{Bakos:2006}. In a multiwavelength set of \textit{HST}/STIS optical transit light curves, \cite{Sing:2011b} observe the typical signature of an occulted spot and, considering the effects of unocculted spots, estimate a spot correction to the transmitted spectrum of about $ 0.00292 \pm 0.00113$ as a fraction of the transit depth in the $320-375$ nm passband, greater than in the $575-625$ nm passband where it is negligible. Moreover, they clearly demonstrate the dependence of the planet-to-star radius ratio on the out-of-transit stellar flux showing that the transit is deeper when the star is fainter as expected as a consequence of unocculted spots present on the disc of the star during the transit (see Sect.~\ref{radius}). In \textit{Spitzer}/IRAC wavebands, both \cite{Beaulieu:2008} and \cite{Desert:2011a} detect variations in the planet-to-star radius ratio. The former authors find that the infrared transit depth is smaller than in the optical \citep[cf. Fig.~4 of][]{Beaulieu:2008} and the effect of stellar spots is to increase this depth. By correcting for this effect, they find that the transit depth is shallower by about 0.19\%, at 3.6 $\mu$m, and 0.18\%, at 5.8 $\mu$m. In the \textit{Spitzer}/IRAC's 3.6 $\mu$m band, the latter authors observe a greater $R_{\rm p}/R_{\rm \star}$ ratio in low brightness periods because of the presence of starspots, implying that the apparent planet-to-star radius ratio varies with stellar brightness  from 0.15566$_{-0.00024}^{+0.00011}$ to 0.1545 $\pm$ 0.0003 \citep{Desert:2009} from low to high brightness periods, respectively.

  A less active M-type star which shows a rotational flux modulation of $\sim$2\% with a period of $\sim$50 days is GJ~1214 \citep{Charbonneau:2009}. Considering the error bars, the planet-to star radius ratio measurements through optical and near-infrared observations \citep{Sada:2010,Berta:2011,Carter:2011} are all consistent with the results of \citealt{Charbonneau:2009} ($0.1162 \pm 0.00067$), obtained in the optical bands. Furthermore \cite{Berta:2011} detect variations in the planet-to-star radius ratio at different epochs and assess the expected variation of this ratio due to stellar variability in optical wavelengths (see their Fig.~8). They state that this effect may be neglected in \textit{Spitzer}/IRAC bands. \cite{Croll:2011} observe a deeper transit in the $K_{\rm S}$ band than in the $J$ band and, after taking into account to effects of the starspots, explain this effect as due to a source of absorption in the  $2.2-2.4$ $\mu$m range, possibly associated with methane \citep{Miller-Ricci:2010}.

  A remarkable study of the starspot effects on the exoplanet sizes is carried out by \cite{Czesla:2009} on the active star CoRoT-2 \citep{Lanza:2009,Huber:2010}. They point out the importance of the normalization of the transit profile to a common reference level to make different transits comparable with each other and show that the relative light loss during a transit is significantly correlated with the out-of-transit flux. The correlation is compatible with a distribution of starspots occulted during the transit that has a larger covering factor when the star is more active, i.e., when the out-of-transit flux is lower. A method to derive the unperturbed transit profile is proposed by extrapolating the observed profiles towards their lower flux limit. A best fit to this profile gives a planet radius about 3\% larger than that of \cite{Alonso:2008}, which is estimated without considering stellar activity in the CoRoT white bandpass ($350-1100$ nm).

  The above examples point out how stellar magnetic activity has a significant impact on the transit depth and in the consequent derivation of the planet radius and density, both in the optical and in the infrared wavebands, thus leading in several cases to contrasting results. As shown, the study of stellar variability is limited to its detection during individual real observations,  without a  systematic theoretical treatment to better characterize its impact through the analysis of stellar colours. This deficiency may lead to create \lq\lq ad hoc\rq\rq ~functional relations for each specific situation. (For instance, the Eq.~(4) in \citet{Carter:2011} represents the slight brightness increasing due to the starspot occultation during a planetary transit and holds only when a circular spot, with the same planetary size, is considered.) A theoretical characterization of transit light curves may help the correct definition of planetary and stellar parameters.

  In this study, we  present a simple approach to better constrain the spot contribution in the optical and infrared transit light curves, thus quantify the flux variations induced by stellar magnetic activity as a function of the wavelength passband, and investigate the starspot effects on the colours of stars. Moreover we present a unified theoretical approach in order to better constrain the impact of both occulted and unocculted starspots on the determination of planetary and stellar parameters. Therefore in the context of the transmission spectroscopy, we examine the dependence of the apparent planetary radius on the magnetic activity level and wavelength.

 \section{Method}
  \label{method}
  Our primary aim is to synthesize the flux variation in solar-type stars when  active regions (dark spots in the considered case) stand on the stellar surface, to understand the effect induced on the light curve of a planetary transit. We focus on the optical and infrared bands, in the Johnson-Cousins-Glass photometric system (UBVRIJHK bands) and in the \textit{Spitzer}/IRAC wavebands centered at 3.6 $\mu$m, 4.0 $\mu$m, 5.8 $\mu$m and 8.0 $\mu$m. 

  We use the standard BaSeL library of synthetic stellar spectra, worked out by \citet{Lejeune:1997,Lejeune:1998} and \citet{Westera:2002}. This library is based on the grids of model atmospheres spectra of \citet{Bessell:1989,Bessell:1991}, \citet{Fluks:1994} and \citet{Kurucz:1995} and covers a wide range of parameters: effective temperature 2000 $\mathrm{K}$ $<$ $T_\mathrm{eff}$ $<$ 50000 $\mathrm{K}$, gravity -1.02 $<$ log \textit{g} $<$ +5.5, metallicity -5.0 $<$ [M/H] $<$ +1.0, wavelength 9.1 nm $<$ $\lambda$ $<$ 160000 nm. The stellar synthetic spectra have a wavelength resolution of 1 nm in the ultraviolet, 2 nm in the visible, $5-10$ nm in the near infrared, $20-40$ nm in the \textit{Spitzer}/IRAC bands.
  \begin{figure}
   \centering
   \includegraphics[width=9.cm]{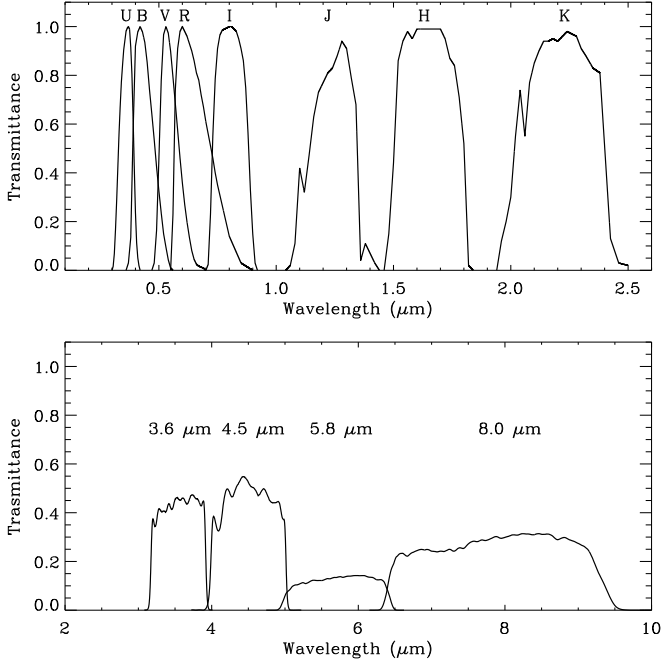}
   \caption{The transmittance of the filter set adopted in this work. \textit{Top}: the Johnson-Cousins-Glass photometric system. \textit{Bottom}: the {\it Spitzer}/IRAC passbands.}
   \label{f:filters}
  \end{figure}

  To derive the stellar flux in the Johnson-Cousins-Glass photometric system and in the {\it Spitzer}/IRAC bands, we convolve the stellar flux distribution derived from the BaSeL libraries with the trasmission curves of the filters shown in Fig.~\ref{f:filters}. We choose as UBVRI filter passbands those adopted by \citet{Bessell:1990}, as JHK filter passbands those adopted by \citet{Bessell:1988}, and the \textit{Spitzer}/IRAC filter passbands as reported in the \textit{Spitzer Science Center} web site\footnote{http://ssc.spitzer.caltech.edu/irac/calibrationfiles/spectralresponse/.}. Both synthetic spectra and filter trasmission curves are interpolated in wavelength at the desired resolution. 

  \subsection{Modelling stellar activity}
   \label{activity}
   We assume a temperature difference between the star and the spot, $\Delta T = T_{\rm \star} - T_{\rm s}$, where $T_{\rm \star}$ and $T_{\rm s}$ represent the effective temperatures of the unperturbed photosphere of the star (e.g. with no stellar activity) and of the spot, respectively. Following the same approach of \citet{Marino:1999}, the measured flux of the system \lq\lq star + spots\rq\rq, $F_\mathrm{\star+s}$, corresponding to the specified temperature difference, is
   \begin{equation}
    F_{\rm \star + s} (\lambda) = (1 - f) F_{\rm \star}(\lambda) + f F_{\rm s} (\lambda),
    \label{eq:star+spot}
   \end{equation}
   where $F_{\rm \star}$ and $F_{\rm s}$ are the unperturbed and the spotted photospheric fluxes, respectively, $f$ is the spot filling factor, or covering factor, which represents the stellar disc fraction covered by spots, and $\lambda$ the mean wavelength of  the passband. The relative variation of the stellar flux produced by starspots with respect to the unperturbed photosphere is:
   \begin{equation}
    \frac{\Delta F}{F} \equiv \frac{F_{\rm \star}(\lambda) - F_ {\rm \star +s}(\lambda) }{F_{\rm \star}(\lambda)} = A_{\lambda} f, 
    \label{eq:def_a}
   \end{equation}
   where the contrast $A_{\lambda} \equiv 1-F_{\rm s}(\lambda)/F_{\rm \star}(\lambda)$ is a function of the passband. Similar approaches may be found in \citet{Desert:2011a,Berta:2011} and \citet{Carter:2011}.

   In the computation of the stellar flux, we assume that the stellar and the spot intensities vary according to the same quadratic limb-darkening law, viz.:
   \begin{equation}
    I(\lambda, \mu) = I(\lambda, 1) [1-u_{1 \lambda}(1-\mu) - u_{2 \lambda} (1-\mu)^{2}],
    \label{eq:limbdarkening} 
   \end{equation}
   where $I(\lambda, \mu)$ is the specific intensity at wavelength $\lambda$, $\mu \equiv \cos \theta$, with $\theta$  the angle between the normal to the stellar surface and the line of sight, and $u_{1 \lambda}$ and $u_{2 \lambda}$ are  the limb-darkening coefficients (LDCs) at the given wavelength. Given the difference in the effective temperature, the LDCs of the unperturbed and the spotted photospheres are actually different, but in the modelling of  the stellar and solar light curves they are assumed to be the same to simplify the treatment \citep[see, e.g.,  ][]{Lanza:2003,Lanza:2004}. In our case, we have a maximum difference between the LDCs of $\sim$30\% in the U passband when considering main-sequence stars with an effective temperature in the range $3700-6000$~K and spots with a  temperature lower by $1250$~K. In view of the advantages of assuming the same limb-darkening profile and the other uncertainties present in the problem, we shall neglect such a difference (cf. Sect.~\ref{radius}). 

   The effect of the wavelength dependence of the limb darkening is shown in the case of CoRoT-2 by \citealt{Czesla:2009} using optical light curves at different wavelengths. The LDCs decrease from the U band to the infrared giving a more flat-bottomed transit at longer wavelengths with shorter ingress and egress profiles \citep[see discussion in Sect.~6 and Fig.~11,][]{Seager:2003,Richardson:2006} that can improve the determination of the transit parameters. Thus even if we consider limb darkening in the derivation of the transit profiles, taking into account that it decreases towards longer wavelengths (see Table 1 in electronic form, \citealt{Claret:1995}), its impact will not be so crucial, since our study is focused primarily on infrared passbands (see \citealt{Deming:2007,Carter:2008,Berta:2011}).

   The stellar microvariability may be  due also to the presence of faculae, bright photospheric regions surrounding spots. They exhibit centre-to-limb variation of their contrast and their brightness depends on size, position on the stellar disc and wavelength. In the case of the Sun, they can produce a relative increase of the  solar irradiance of $\sim$10$^{-3}$ \citep{Froehlich:2004} at the maximum of the 11-yr cycle. They are visible primarily in white light and the contrast to the unperturbed photosphere is reduced in the near-infrared \citep{Solanki:1998,Froehlich:2004}, since it depends on $\lambda^{-1}$ \citep{Chapman:1977}. In the present study we neglected their contribution since we are interested in the infrared bands where the facular effect may be considered less important. 
   Moreover, the facular contribution to the optical flux variations is negligible in the case of stars significantly more active than the Sun \citep[cf., e.g., ][ and references therein]{Lanza:2009}.

  \subsection{Effects of starspots on the apparent planetary radius}
   \label{radius}
   Starspots  affect in two ways the stellar flux variations in a transit light curve. Spots that are not occulted by the planet  produce a decrease of the out-of-transit flux of the star $F_{\rm OOT}$ that is the reference level to normalize the transit profile. Spots that are occulted during the transit produce a relative increase of the flux because the flux blocked by the planetary disc is lower than in the case of the unperturbed photosphere. The occultation of a starspot is therefore associated with a bump in the transit profile whose duration depends on the spot size and whose height depends on the spot contrast \citep[e. g., ][]{Pont:2007,Sing:2011b}. If we denote the flux during the transit as $F_{\rm IT}$, we have:
   \begin{equation}
    F_{\rm OOT}(\lambda) = F_{\rm \star}(\lambda) - \Delta F_{\rm s}(\lambda)
    \label{eq:OOT_un},
   \end{equation}
and
   \begin{equation}
    F_{\rm IT}(\lambda, \mu) = F_{\rm OOT}(\lambda) - \langle I(\lambda, \mu)\rangle ~\pi R_{\rm p}^{2},
    \label{eq:IT_un}
   \end{equation}
   where $\Delta F_{\rm s} = A_{\lambda} f_{0} F_{\rm \star}$ is the flux decrease due to the unocculted spots having a filling factor $f_{0}$ (cf. Eq.~\ref{eq:def_a}), $\langle I_{\rm \star}(\lambda, \mu)\rangle $ is the specific intensity averaged over the portion of the stellar disc occulted by the planet of radius $R_{\rm p}$. If we denote with $f_{\rm i}$ the filling factor of the spots occulted during the transit when the planet's centre is at the disc position $\mu$ along the transit chord, then 
   \begin{equation}
    \langle I(\lambda, \mu)\rangle = (1- f_{\rm i}) \langle I_{\rm \star} (\lambda, \mu) \rangle  + f_{\rm i} \langle I_{\rm s}  (\lambda, \mu) \rangle = \langle I_{\rm \star}  (\lambda, \mu) \rangle (1-f_{\rm i}A_{\lambda}),
    \label{spec_intens}
   \end{equation}
   where $\langle I_{\star}(\lambda, \mu) \rangle$ is the average of the specific intensity of the unperturbed photosphere over the area occulted by the planet and $\langle I_{\rm s}  (\lambda, \mu) \rangle$ is the same quantity for the spotted photosphere. Note that Eq.~(\ref{spec_intens}) is valid because  we have assumed that the spotted and unspotted photospheres have the same LDCs. If we define the relative transit profile at wavelength $\lambda$ as $D(\lambda, \mu) \equiv 1 - F_{\rm IT} (\lambda, \mu)/F_{\rm OOT}(\lambda)$, after some simple algebra we find:
   \begin{equation}
    D(\lambda, \mu) = \left(\frac{R_{\rm p}}{R_{\rm \star}} \right)^{2} G(\mu) ~\frac{1-f_{\rm i}A_{\lambda }}{1-f_{0}A_{\lambda}}, 
    \label{eq:deficit_oc}
   \end{equation}
   where $R_{\star}$ is the radius of the star and $G(\mu)$ is a function of the limb-darkening coefficients and the disc position $\mu$ of the planet during the transit, but does not depend on the spot distribution and temperature. If we define the unperturbed transit profile as: $D_{0} (\lambda ,\mu) \equiv  \left(\frac{R_{\rm p}}{R_{\rm \star}} \right)^{2} G(\mu)$, the profile in the presence of spots is:
   \begin{equation}
    D(\lambda, \mu) = D_{0}(\lambda, \mu) ~\frac{1-f_{\rm i}A_{\lambda }}{1-f_{0}A_{\lambda}}, 
    \label{xx}
   \end{equation}
   If the ratio  $R_{\rm p} / R_{\star} \la 0.1$, we can assume that $\langle I_{\rm \star}(\lambda, \mu)\rangle \simeq I_{\star} (\lambda, \mu)$ \citep{Mandel:2002} and the expression of the function $G(\mu) $ becomes particularly simple:
   \begin{equation}
    G(\mu) = \frac{1 - u_{1 \lambda}(1-\mu) - u_{2 \lambda}(1-\mu)^{2}}{1 - \frac{u_{1 \lambda}}{3} - \frac{u_{2 \lambda}}{6}}.
    \label{eq:G}
   \end{equation}
   In Eq.~(\ref{eq:deficit_oc}) the filling factor of the occulted spots $f_{\rm i}$ is in general a function of the position of the centre of the planetary disc along the transit chord, while the filling factor of the out-of-transit spots $f_{0}$ can be assumed to be constant because the visibility of those spots is modulated on timescales much longer than that of the transit, i.e., of the order of the stellar rotation period that is generally of several days. Moreover, when a geometric model is fitted to an observed transit profile, the bumps produced by occulted spots are excised from the data set to reduce their impact as much as possible. However, the capability of detecting the flux bump due to a particular spot depends on its size and contrast as well as on the accuracy and cadence of the photometry.  If there is a more or less continuous background of spots along the transit chord, they cannot be individually resolved and the transit profile will appear to be symmetric and without any obvious indication of starspot perturbations. This is indeed the most dangerous case when fitting an observed transit profile. In this case, $f_{\rm i}$ can be assumed to be constant during a given transit because otherwise we would have detected the individual spot bumps. If we assume that both $f_{0}$ and $f_{\rm i}$ are much less than unity, we can develop  the denominator of Eq.~(\ref{eq:deficit_oc}) in series and neglecting the second order terms in the filling factors, we find:
   \begin{equation}
    D(\lambda, \mu) \simeq  \left(\frac{R_{\rm p}}{R_{\rm \star}} \right)^{2} G(\mu) \left[ 1 - A_{\lambda}(f_{\rm i} - f_{0}) \right]. 
    \label{eq:deficit_oc1}
   \end{equation}
   From Eq.~(\ref{eq:deficit_oc1}), we see that the effects of the unocculted and occulted spots tend to compensate each other although the filling factor of the occulted spots may generally be higher because starspots tend to appear at low or intermediate latitudes in Sun-like stars. 
   \begin{table*}[!htbp]\footnotesize
\caption{Examined cases and their related real planetary systems. In the synthesized models the metallicity and gravity values are the solar ones.}
\label{t:cases}
\centering
\begin{tabular}{c|cccc|ccccc}
\hline
\hline
\multirow{3}*{Case} & \multicolumn{4}{c}{\textit{Reference Systems}} & \multicolumn{5}{|c}{\textit{Real Systems}} \\
\cline{2-10}
& Spectral & T$_{*}$ & T$_{s}$ & $\Delta$T & Star &  Spectral & R$_{p}$ & M$_{p}$ & $\Delta$m \\
& Type & (K) & (K) & (K) & name & Type  & (R$_{J}$) & (M$_{J}$) & \\
\hline
1) & Early-G & 5750 & 4500 & 1250 & CoRoT-12 & G2V  & 1.44 $\pm$ 0.13 & 0.917 & 0.018\\
2) & Early-K & 5250 & 4000 & 1250 & CoRoT-7 & K0V  & 0.15 $\pm$ 0.008 & 0.0151 & 3.14$\cdot$10$^{-4}$\\
3) & Mid-K & 4500 & 3250 & 1250 & HAT-P-20 & K7  & 0.867 $\pm$ 0.033 & 7.246 & 0.016\\
4) & Early-M & 3750 & 2500 & 1250 & GJ 436 & M2.5  & 0.365 $\pm$ 0.02 & 0.0737 & 6.53$\cdot$10$^{-3}$\\
\hline

\end{tabular}
\end{table*}

   \begin{table*}
\caption{Contrast coefficients A$_{\lambda}$ for each reference system in the different passbands.}
\centering
\begin{tabular}{c|cccccccccccc}
\hline
\hline
Case & A$_{U}$ & A$_{B}$ & A$_{V}$ & A$_{R}$ & A$_{I}$ & A$_{J}$  & A$_{H}$ & A$_{K}$ & A$_{3.6}$ & A$_{4.5}$ & A$_{5.8}$ & A$_{8.0}$ \\
\hline
1) & 0.938 & 0.854 & 0.759 & 0.686 & 0.616 & 0.458 & 0.316 & 0.285 & 0.248 & 0.283 & 0.262 & 0.232\\
2) & 0.958 & 0.901 & 0.833 & 0.756 & 0.643 & 0.560 & 0.456 & 0.429 & 0.392 & 0.397 & 0.373 & 0.348\\
3) & 0.960 & 0.959 & 0.946 & 0.912 & 0.779 & 0.614 & 0.605 & 0.547 & 0.495 & 0.444 & 0.496 & 0.503\\
4) & 0.994 & 0.995 & 0.992 & 0.978 & 0.924 & 0.722 & 0.733 & 0.640 & 0.542 & 0.543 & 0.656 & 0.677\\
\hline
\end{tabular}
\label{t:A_lambda}
\end{table*}

   An alternative interpretation of Eqs.~(\ref{xx}) and (\ref{eq:deficit_oc1}) is that spots change the apparent relative radius of the planet $(R_{\rm p}/R_{\star})_{\rm a}$ as derived from the fitting of the transit profile (see Sect.~\ref{param_var}) as:
   \begin{equation}
    \left( \frac{R_{\rm p}}{R_{\star}}  \right)_{\rm a} = ~\left( \frac{R_{\rm p}}{R_{\star}}  \right) \sqrt{\frac{1-f_{\rm i}A_{\lambda }}{1-f_{0}A_{\lambda}}}  \simeq \left( \frac{R_{\rm p}}{R_{\star}}  \right) \left[ 1  - \frac{1}{2} A_{\lambda} (f_{i} - f_{0}) \right]. 
   \end{equation}
   Therefore, the apparent variation of the planetary radius is:
   \begin{equation}
    \frac{\Delta R_{\rm p}(\lambda)}{R_{\rm p}}   \simeq - \frac{1}{2} A_{\lambda} (f_{\rm i} - f_{0}),  
    \label{eq:depth2}
   \end{equation}
   When radius measurements are available in two different passbands $\lambda_{1}$ and $\lambda_{2}$, we have:
   \begin{equation}
    \left( \frac{R_{\rm p}}{R_{\star}}  \right)_{\rm a} (\lambda_{1}) - \left( \frac{R_{\rm p}}{R_{\star}}  \right)_{\rm a} (\lambda_{2}) \simeq - \frac{1}{2} (A_{\lambda_{1}} - A_{\lambda_{2}})  (f_{\rm i} - f_{0}).
    \label{eq:rvsl}
   \end{equation}
   If the spot temperature is known, this equation can be used to derive the effective filling factor $ f \equiv f_{\rm i} - f_{0}$ from  the apparent radius measurements in two well-separated passbands, e.g., in the U and the K passbands. If the spot temperature is not known, it can be derived when a third measurement is available in a further passband, thereby removing the degeneracy between filling factors and contrast. However, the presence of a planetary atmosphere with a wavelength-dependent  opacity can complicate the problem as we shall see in the final part of Sect.~\ref{depth} in the case of HD~189733.

  \subsection{Effects of starspots on the determination of the orbital parameters}
   \label{param_var}
   The perturbations of the transit profile due to starspots do not affect only the determination of the relative radius of the planet $R_{\rm p}/R_{\star}$ but also the other geometrical parameters of the system and the limb-darkening coefficients. Most of the previous studies have considered only the radius variation \citep[cf., e. g., ][]{Berta:2011,Carter:2011}, but the impact on the other parameters can be significant. \citet{Czesla:2009} explored the variation of the planetary radius and orbital inclination keeping fixed the semimajor axis of the orbit $a$, the stellar radius $R_{\star}$, and the limb-darkening coefficients $u_{1}$ and $u_{2}$ at the values of \citet{Alonso:2008}. In this way, they found that the radius of the planet is increased by $\sim$3\% by fitting the lower envelope of the transit profiles derived with their method. On the other hand, they found no significant correction in the inclination $i$ of the orbital plane of the planet to the plane of the sky. 

   A limitation of the approach by \citet{Czesla:2009} is that by fixing the ratio $a/R_{\star}$, the duration of the transit $t_{\rm T}$ fixes the inclination because $t_{\rm T} =  \sqrt{1 - b^2} P_{\rm orb}/  (\pi a/R_{\star})$, where $b =  a/R_{\star} \cos i$ is the impact parameter and $P_{\rm orb}$ the orbital period. Since the effects of the spots on the duration of the transit are generally small, the inclination is not changed by fitting their unperturbed transit profile. Therefore, to perform a proper estimate of the systematic effects produced by occulted and unocculted spots, we cannot fix any of the system parameters in fitting the transit profile. 

   The unperturbed transit profile $D_{0} (\lambda ,\mu)$ can be derived from the lower envelope of the observed transit profiles if we can guess the minimum spot filling factor of the unocculted spots $f_{\rm 0 \,\, min}$ and of the spots occulted during transits $f_{\rm i \,\,  min}$. From the lower envelope of the observed transits $D_{\rm low}(\lambda, \mu)$, we get an estimate of the unperturbed profile as:
   \begin{equation}
    D_{0} (\lambda, \mu) = D_{\rm low} (\lambda, \mu) \frac{1-A_{\lambda}f _{\rm 0 \,\, min}}{1- A_{\lambda} f_{\rm i \, \, min}}.
   \end{equation} 
   Since $f_{\rm i \, \, min}$ is a constant, the unperturbed profile does not show any  bumps due to the occultation of individual spots and is symmetric with respect to the mid of the transit. 

   We fit the unperturbed  profile  using an analytical model for the transit, e.g., that of \citet{Pal:2008} that has the advantage of providing analytic expressions for the derivatives of the transit profile $D_{0}$ with respect to the transit parameters allowing us an efficient application of the Levenberg-Marquardt method to minimize the $\chi^2$ \citep{Press:1992}. To minimize their mutual correlations, we choose as free parameters for our best fitting: $\zeta / R_{\star} \equiv (2\pi / P_{\rm orb}) (a/R_{\star}) / \sqrt{1 - b^2}$, $b^2$, $R_{\rm p}/R_{\star}$,  and the limb darkening parameters $u_{+} \equiv u_{1} + u_{2}$ and $u_{-} \equiv u_{1} - u_{2}$ \citep[see ][ for a justification of the independence of this parameter set]{Pal:2008,Alonso:2008}. The orbital period and the mid transit epoch are kept fixed in the fitting because they can be derived from the observation of a sufficiently long sequence of transits. Specifically, the profile distortions due to resolved occulted spots may affect the timing of  individual transits, but the spot perturbations  can be assumed to be random so that they increase  the statistical error of the timing. Therefore, a sufficiently long time series can be used to derive an accurate value of $P_{\rm orb}$. We shall present an application of this method to estimate the impact of starspots on the fitting of the parameters of the system CoRoT-2 in Sect.~\ref{star_parameter}.

 \section{Results}
  \label{results}
  We focus our analysis on star temperature values ranging from the solar one ($\sim$5750~$\mathrm{K}$) to that of an early M-type star and set the temperature difference between the star and the spot to 1250~$\mathrm{K}$ (see \citealt{Berdyugina:2005}). The synthetic  spectra, both for the unperturbed photosphere and the spots, were chosen for solar metallicity  and gravity, i.e. $[M/H]_\mathrm{\odot}$ = 0.0 and log $g_\mathrm{\odot}$ = 4.44.  To apply our results to real cases, we selected four systems with planets, i.e.,
  CoRoT-12, with a transiting Jupiter-like planet orbiting around a G2V star \citep[see][]{Gillon:2010}; CoRoT-7, a K0V star  \citep[][]{Leger:2009,Queloz:2009}; HAT-P-20, a K7V star  \citep[][]{Bakos:2010}; and GJ 436, a M2.5 dwarf star harbouring a transiting Neptune-mass planet \citep[][]{Butler:2004}. 
   \begin{figure}[!htbp]
    \centering
    \resizebox{\hsize}{!}{\includegraphics[width=9.cm]{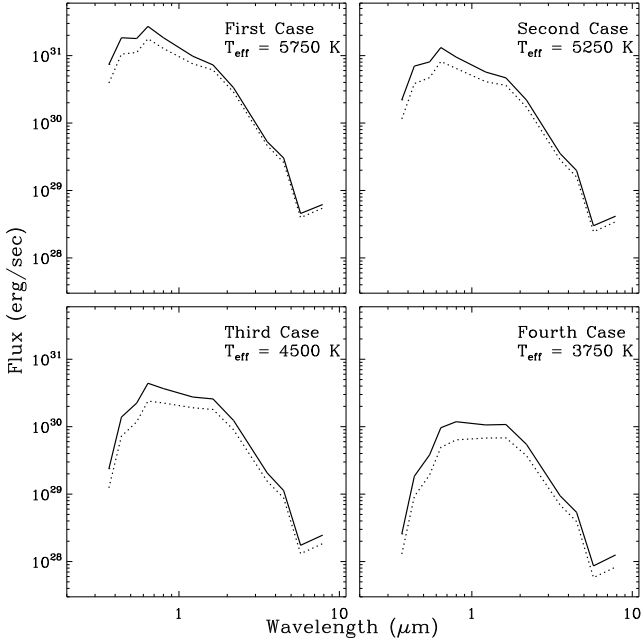}}
    \caption{\small Flux vs. Wavelength for all the cases in Table~\ref{t:cases}.  Solid lines plot the spectral distributions of the stellar unperturbed photospheres and dotted lines the spectral distributions of the star + spot systems with \textit{f} = 0.5. In the \textit{top left} we present the first case, in the \textit{top right} the second case, in the \textit{bottom left} the third case and in the \textit{bottom right} the fourth case.}
    \label{f:flux_vs_lambda}
   \end{figure}
   We list their properties in Table~\ref{t:cases}, where we give, from the left to the right, the spectral type, the selected temperature for the quiet photosphere and spot for the reference stars, the difference between photosphere and spot temperatures, the real system with the name of the star we will use for comparison, the stellar spectral type, the planetary radius and mass, and the reduction in flux at mid planetary transit, expressed in magnitudes (obtained as the square of the ratio of the planet radius to the star radius), respectively. 

  The contrast coefficients $A_{\rm \lambda}$, giving the spot-induced flux variations for each of the examined cases (see Sect.~\ref{activity}), are obtained by choosing the spot filling factor of the out-of-transit spots, $f_{0}$, as ranging from 0.01 to 0.5 of the stellar disc surface \citep{Berdyugina:2005} and are reported in Table~\ref{t:A_lambda}. For G and early-K  stars, the minimum flux perturbation occurs in the 8.0~$\mu$m {\it Spitzer}/IRAC passband, while the maximum occurs in the U passband because the spot contrast increases monotonically towards shorter wavelengths. In the case of late-K and M stars, the minimum flux perturbation occurs in the 4.5~$\mu$m and 3.6~$\mu$m {\it Spitzer}/IRAC passband, respectively, due to molecular bands in the spectrum. These features are observed in the spectra of cooler stars, both in the optical and in the infrared, and this results in a non-monotonous spectral variation in correspondence of these bands. 
  In this case, the unperturbed and spotted spectra do not behave just as power laws in the infrared domain as in the hotter cases.

   \begin{figure}[!htbp]
    \centering
    \resizebox{\hsize}{!}{\includegraphics[width=9.cm]{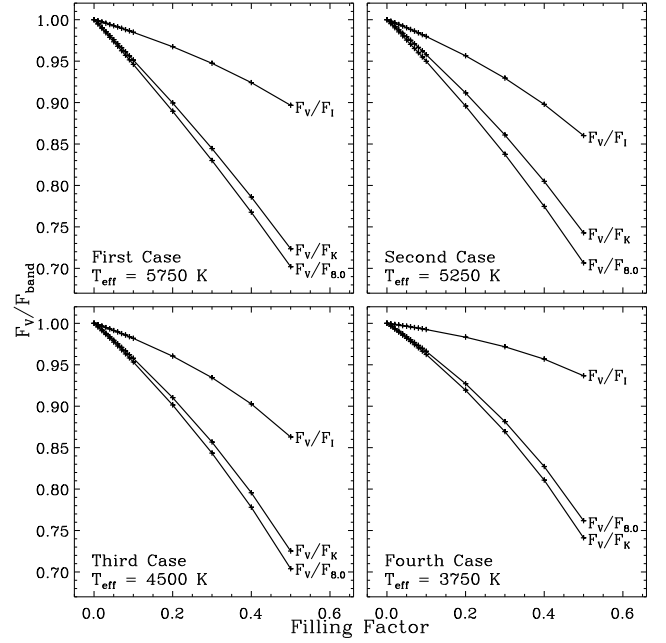}}
    \caption{\small Normalized $F_\mathrm{V}$/$F_\mathrm{band}$ vs. Filling Factor $f_{0}$ for all the cases in Table~\ref{t:cases} in different passbands: I, K and 8.0 $\mu$m. Different curves plot different ratios: $F_\mathrm{V}$/$F_\mathrm{I}$, $F_\mathrm{V}$/$F_\mathrm{K}$ and $F_\mathrm{V}$/$F_\mathrm{8.0}$, as labelled. In the \textit{top left} we present the first case, in the \textit{top right} the second case, in the \textit{bottom left} the third case, and in the \textit{bottom right} the fourth case.}
    \label{f:colours}
   \end{figure}
  In Fig.~\ref{f:flux_vs_lambda} we plot the system flux vs. wavelength for all the cases, for a given filling factor of the out-of-transit spots. The solid curve represents the spectrum of the unperturbed photosphere and the dotted line the spectral distribution of the system \lq\lq star + spots\rq\rq. We have plotted a case with an extreme filling factor ($f_{0} = 0.5$) to make the effect of the spots clearly evident. The spot perturbation decreases with increasing wavelength,  although the details of the variation depends on the stellar effective temperature. 

  In Fig.~\ref{f:colours} we plot the variations of the normalized ratios $F_\mathrm{V}$/$F_\mathrm{band}$ vs. the filling factor $f_{0}$, where $F_\mathrm{band}$ is computed in three specified passbands (I, K and 8.0 $\mu$m), with the purpose of understanding to what extent the stellar magnetic activity may produce a variation in the colours of the star (this graph will be discussed in details in sect.~\ref{colours}).

  These results suggest that the colour effect due to stellar activity has to be taken into account if we analyse the photometric signals even of a moderately active star (comparable with or slightly more active than our Sun). The effects are both on stellar colours (discussed in  Sect.~\ref{colours}) and on the determination of the radius of the planet (discussed in Sect.~\ref{depth}). 

  \subsection{Influence of stellar activity on stellar colours and the characterization of a single light curve dip}
   \label{colours}
   If we consider the first and the fourth cases in Table~\ref{t:cases} (i.e., the hottest and the coolest host stars), by fixing the filling factor $f_{0}$ to 1\% (i.e., larger than in the solar case, but comparable to those observed in late-type moderately active stars), we can compute the relative variation of the system flux ($\Delta F/F$) in the U and K-bands,  to compare the influence of stellar activity in the optical and  infrared passbands.

   For the first case ($T_{\rm \star}$ = 5750 $\mathrm{K}$, $\Delta T$ = 1250 $\mathrm{K}$) we derive the following variations in the U-band and in the K-band, respectively:
   \begin{equation}
    \Delta F_{U}/F_{U} = 9 \cdot10^{-3} \nonumber \\
    $and$\\
    \Delta F_{K}/F_{K} = 3 \cdot10^{-3}. \nonumber
    \label{eq:first}
   \end{equation}

   For the fourth case ($T_{\rm \star}$ = 3750 $\mathrm{K}$, $\Delta T$ = 1250 $\mathrm{K}$) we derive the following values in the U-band and in the the K-band, respectively:
   \begin{equation}
    \Delta F_{U}/F_{U} = 1 \cdot10^{-2} \nonumber \\
    $and$\\
    \Delta F_{K}/F_{K} = 6 \cdot10^{-3}. \nonumber
    \label{eq:fourth}
   \end{equation}

   The decrease in flux caused by a spot filling factor of 0.01 is comparable with the decrease produced by the transit of a Jupiter-sized planet in front of the stellar disc. In fact we can compare the obtained reductions in flux of the system due to stellar magnetic activity,  using the information in Table~\ref{t:A_lambda}, with the decrease in magnitude due to the transit of a planet, reported in Table~\ref{t:cases}.

   Our results indicate that it is easier to identify the transit of a Jupiter-like planet by analysing the light curve of the system in the K-band, while it may be more problematic in the U-band.
   The detection of a smaller planet is more challenging, both in the U and in the K-band, since the photometric signal corresponding to the transit of such a planet is, as expected, smaller than the effect due to the magnetic activity of its parent star. Fortunately other properties of the variations, as timescales or colour dependence, may help in understanding the origin of the variations.

   The transit of a spot or a group of spots (or an active region in general) on the star surface  induces a  colour variation of the star. Since spots are cooler than the stellar photosphere, the  colour indexes are increased. In principle, we can infer the activity of a star by monitoring the variation in the colour  for some specified wavebands. Star colours depend on the ratio between the fluxes in two selected passbands. For this purpose, we compute the ratio between the system fluxes in the optical and the infrared passbands, for all the four template  cases: $F_\mathrm{V}$/$F_\mathrm{I}$, $F_\mathrm{V}$/$F_\mathrm{K}$ and $F_\mathrm{V}$/$F_\mathrm{8.0}$. In Table~\ref{t:tab_colours} we report the variation of $F_\mathrm{V}$/$F_\mathrm{band}$ in the different bands, by fixing the filling factor $f_{0}$ to 0.1 and to 0.01, respectively. The first column shows the filling factors and the enumeration of the cases, the following columns show the changes in $F_\mathrm{V}$/$F_\mathrm{I}$, in $F_\mathrm{V}$/$F_\mathrm{K}$, and in $F_\mathrm{V}$/$F_\mathrm{8.0}$, respectively. The behaviour of the star flux ratios vs. spot filling factor of the unocculted spots are plotted in Fig.~\ref{f:colours}, where the ratios decrease as the spot filling factor increases.

   The transit of a planet without atmosphere induces a relative decrease of the stellar flux  that depends on the mean wavelength of the passband owing to the wavelength dependence of the stellar limb darkening. For the present application, we consider the central transit ($b=0$) of a small planet $(R_{\rm p}/R_{\star}) \la 0.1$, with the quadratic limb-darkening law of Eq. (\ref{eq:limbdarkening}). We derive the ratio $\cal{R}$ of the relative depths of the transit at the mean wavelengths of two different passbands, say U and K,
   \begin{equation}
    {\cal R} \equiv \frac{\Delta F_{U}/F_{U}}{\Delta F_{K}/F_{K}} = \frac{1-\frac{1}{3}u_{1 K} -\frac{1}{6}u_{2 K} }{1-\frac{1}{3} u_{1 U} - \frac{1}{6} u_{2 U} }, 
    \label{eq:r_limb}
   \end{equation}
   where $u_{1 U}$, $u_{2 U}$, $u_{1 K}$, and $u_{2 K}$ are the limb-darkening coefficients in the U and K passbands, derived from, e. g.,  \cite{Diaz-Cordoves:1995} and \cite{Claret:1995}, respectively. For the first case, we find ${\cal R} = 1.298$, while for the fourth case, ${\cal R}=1.197$. This is significantly smaller than expected in the case of the variations induced by dark spots 
   (cf. (\ref{eq:first}) and (\ref{eq:fourth})). The flux in the K passband coming from the night side of the planet during the transit is neglected because the planet temperature is always lower than $\sim$2000~K, even for the strongest irradiated objects, that is remarkably lower than the temperature of the starspots.
   \begin{table}[!ht]\footnotesize
\centering
\caption{Variations in the colour of the system for all the cases. \textit{Top}: computed for \textit{$f_{0}$} = 0.1. \textit{Bottom}: computed for \textit{$f_{0}$} = 0.01.}
\begin{tabular}{c|ccc}
\hline \hline
$f_{0}$ = 0.1 & $\Delta$(F$_{V}$/F$_{I}$) & $\Delta$(F$_{V}$/F$_{K}$) & $\Delta$(F$_{V}$/F$_{8.0}$)\\
\hline
1) & 1.5\% & 4.9\% & 5.4\% \\
2) & 2.0\% & 4.2\% & 5.0\% \\
3) & 1.8\% & 4.2\% & 4.7\% \\
4) & 0.7\% & 3.8\% & 3.4\% \\
\hline
$f_{0}$ = 0.01 & $\Delta$(F$_{V}$/F$_{I}$) & $\Delta$(F$_{V}$/F$_{K}$) & $\Delta$(F$_{V}$/F$_{8.0}$)\\
\hline
1) & 0.14\% & 0.48\% & 0.53\% \\
2) & 0.19\% & 0.41\% & 0.49\% \\
3) & 0.17\% & 0.40\% & 0.45\% \\
4) & 0.07\% & 0.35\% & 0.32\% \\
\hline
\end{tabular}
\label{t:tab_colours}
\end{table}

   Considering Eq.~(\ref{eq:def_a}), we find that the ratio of the variations induced by starspots is:
   \begin{equation}
    {\cal R}_{\rm s} = \frac{A_{U}}{A_{K}}. 
    \label{eq:r_a}
   \end{equation}
   For a stellar effective temperature of 5750 $\mathrm{K}$, this is always significantly greater than $1.3$, except for very cool spots, i.e., $\Delta T$ $\geq$ 2100 $\mathrm{K}$. Indeed, spots so cool are generally not observed in active stars \citep[see, e.g.,][]{Berdyugina:2005}. For the Sun as a star, $\Delta T$ $\simeq 500-600$ $\mathrm{K}$ because sunspot irradiance is dominated by the penumbral regions \citep[cf., e.g., ][ and references therein]{Lanza:2004}. 
   In the case of the star with $T_{\rm \star}$ = 3750 $\mathrm{K}$, a spot temperature deficit  $\Delta T$ $\geq$ 1500 $\mathrm{K}$ is required to have $\cal R_{\rm s}$ comparable with that of a planetary transit, which is also unlikely for such a cool star \citep[see, e.g.,][]{Berdyugina:2005,Zboril:2003}.

   In conclusion, even for an individual  light curve dip, we are able to compare the variation of the depths in the optical and infrared passbands to discriminate between an effect induced by spot activity and a planetary transit, because the ratio $\cal R$ is in the range  $1.7-3.0$ in the former case (see the ratio of relations in (\ref{eq:first}) and in (\ref{eq:fourth}), for a typical $\Delta T$ = 1250 $\mathrm{K}$), while it is around $1.2-1.3$ in the latter case. 

   The timescales of the stellar flux variability and of the transit  are generally quite different, since the former may range from weeks to months, as in the case of the Sun \citep{Lanza:2003,Lanza:2007}, while the latter has a duration ranging from a few hours to a few tens of hours.  However, there are situations in which the timescales can be comparable, e.g., in the case of a young sun-like star with a rotation period of, say, $1-2$ days transited by a planet at the distance of $\sim$1 AU, whose transit duration is of the order of $\sim10-15$ hours. In this case, a discrimination of the origin of the light dip based on timescales is unfeasible while our approach is still applicable if simultaneous multiwavelength observations are available. Note that our method can be applied in principle even to a single event. This can be very useful in the case of planets with orbital periods of several tens or hundreds of days.

   The photometric precision currently achievable from the ground in the near infrared is generally sufficient for an application of the present method,  provided that the  photometric variations are at least  of $0.005-0.01$ mag. \citet{Croll:2011} acquired photometry with a standard deviation of $\sigma \simeq 5 \times 10^{-3}$ mag in the J (1.25 $\mu$m) and the K (2.15 $\mu$m) bandpasses using the WIRCam at the 3.5-m CFHT, while \citet{Tofflemire:2011} reached a photometric stability of $3.9 \times 10^{-3}$ mag in one hour and of $5.1 \times 10^{-3}$ mag in one night in the K$_{\rm s}$ passband at 2.224 $\mu$m observing stable M-type stars. However, the situation is remarkably different in the mid and far infrared passbands.

   From space, the \textit{Warm Spitzer} mission  has reached a statistical error of 450 ppm (parts per million) per minute, observing GJ 1214 for 12 sec at 3.6 $\mu$m \citep{Gillon:2011}. However, the IR observations are affected by systematic effects that may compromise the measurements, since they induce flux variations comparable to the depth of a planetary transit. For example, the 3.6 and 4.5 $\mu$m \textit{Spitzer}/IRAC channels (InSb detectors) are affected by the pixel-phase effect \citep{Reach:2005,Charbonneau:2005,Morales:2006,Knutson:2008}, due to the telescope jitter and intra-pixel variation of the sensitivity of the detector, which may cause a flux peak-to-peak amplitude of up to $\sim$1\% \citep{Beaulieu:2008,Hebrard:2010} or even higher \citep{Desert:2011b,Fressin:2011}, according to the exposure time.
   \begin{figure}[!htbp]
    \centering
    \resizebox{\hsize}{!}{\includegraphics[width=9.cm]{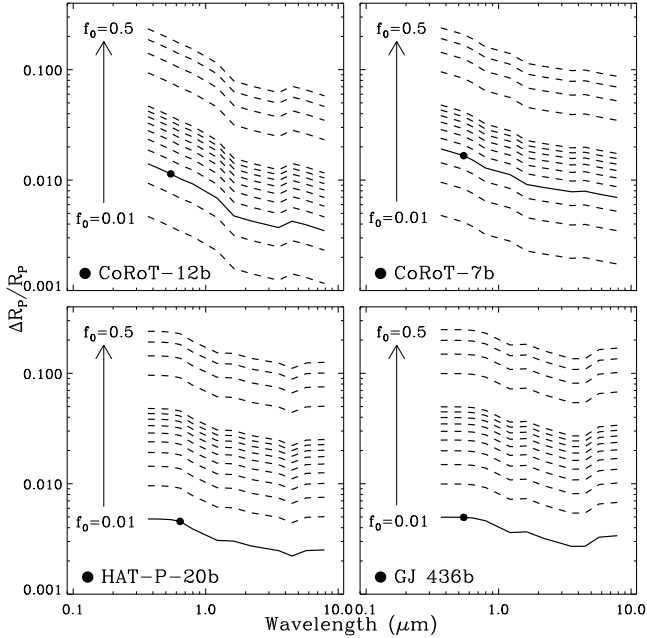}}
    \caption{\small Relative Variation of the apparent planetary radius vs. wavelength for all the examined planetary systems. In the \textit{top left} we present the results  for CoRoT-12b, in the \textit{top right} for CoRoT-7b, in the \textit{bottom left} for HAT-P-20b, and in the \textit{bottom right} for GJ~436b, assuming various levels of activity. The dashed curves, from the bottom to the top, correspond to increasing filling factor values, as indicated on the left of each plot. The solid curves refer to the activity levels as actually estimated for the specified systems. The filled points indicate the position of the observed planets, as described in the text.}
    \label{f:depth}
   \end{figure} 
   The light curves of the 5.8 and 8.0 $\mu$m channels (Si:As detectors) show non-linear  trends vs. time, called the ramp effect \citep{Deming:2006,Knutson:2007b}, caused by the trapping of electrons by the detector impurities (see \citealt{Agol:2010} for further details), that produce mmag level flux variations in photometry. They can reach up to $\sim$10\% for the lowest illuminated pixels over 33 hr \citep{Agol:2010}, and increase over time \citep{Machalek:2008}. The correction of these instrumental systematics and artifacts is particularly relevant for future infrared missions, such as JWST, since the same Si:As technology, present in the last two \textit{Spitzer}/IRAC channels, will be adopted for MIRI (Mid-Infrared Instrument, see \citealt{Gardner:2006}).

   Anyhow, in this study we intend to emphasize the importance of a broad band monitoring of stars hosting planets from visible, at the shortest wavelength as possible, to IR. Simultaneous observations, from optical to infrared wavebands, allow us to better understand the role of stellar magnetic activity in the analysis of a transiting planet light curve comparing the ratios $\cal R$ derived in eqs. (\ref{eq:r_limb}) and (\ref{eq:r_a}), as stated before.

   \subsection{Influence of stellar activity on planetary radius determination}
    \label{depth}
    In this section, we apply the simple formulae of Sect.~\ref{radius} to evaluate the variation of the apparent planetary radius in the presence of stellar activity assuming that the other parameters are not affected. The effect of the activity on the simultaneous determination of the radius and the orbital parameters will be illustrated in Sect.~\ref{star_parameter}.
    \begin{table}[!ht]\footnotesize
\caption{The relative variation in planetary radius computed at optical wavelength for CoRoT-12b, CoRoT-7b e GJ~436b, and in the R-band for HAT-P-20b, taking into account the adopted filling factor values reported in the text. $\Delta$Err is the ratio of the computed $\Delta R_{\rm p}$ to the error bars quoted in the literature.}
\label{t:depth}
\centering
\begin{tabular}{ccc}
\hline \hline
Planet Name & $\Delta$R$_{p}$/R$_{p}$ & $\Delta$Err \\
\hline
CoRoT-12b & 1.14 $\cdot$ 10$^{-2}$ & 13\% \\
CoRoT-7b  & 1.67 $\cdot$ 10$^{-2}$ & 31\% \\
HAT-P-20b & 0.46 $\cdot$ 10$^{-2}$ & 12\% \\
GJ 436b   & 0.50 $\cdot$ 10$^{-2}$ & 10\% \\
\hline
\end{tabular}
\end{table}

    The plots in Fig.~\ref{f:depth}, $\Delta R_{\rm p}$/$R_{\rm p}$ vs. wavelength, give us the overestimate of $R_{\rm p}$ if the effect of unocculted starspots is neglected for our four reference cases. We find a relation between $\Delta R_{\rm p}$/$R_{\rm p}$ and the wavelength and between $\Delta R_{\rm p}$/$R_{\rm p}$ and the stellar activity, i.e. increasing the filling factor of the out-of-transit spots $f_{0}$ the dashed curves shift upwards and the variation decreases towards longer wavelengths. The filled circles on the plot indicate the correction on the  observed planetary radius that we have to apply to get the true radius of the planet for the assumed spot temperature and filling factors (see below).

    In the top left panel of Fig.~\ref{f:depth} we present the result for CoRoT-12. We assumed a filling factor of $f_{0} = 0.03$ (the solid line in the panel) by inferring the flux variation from the CoRoT light curve of Fig.~1 in \citet{Gillon:2010} and then derived the contrast coefficients from the first row of Table~\ref{t:A_lambda}. 

    The top right panel of Fig.~\ref{f:depth} shows $\Delta R_{\rm p}$/$R_{\rm p}$ vs. wavelength for CoRoT-7. The spot filling factor value is $f_{0} = 0.04$ by deducing it from the optical flux variability reported in \citet{Lanza:2010}. 

    For HAT-P-20, the outcomes are shown in the bottom left panel of Fig.~\ref{f:depth}. We set $f_{0}$ = 0.01 as derived by the magnitude variations of the stellar flux in the R-band shown in Fig.~1 in \citet{Bakos:2010}. The dependencies of $\Delta R_{\rm p}$/$R_{\rm p}$ on the wavelength and the stellar activity are clearly visible for this planet too, even though the curves tend to become flat.

    The bottom right panel of Fig.~\ref{f:depth} shows the result for GJ~436. The star is a relatively quiet source showing photometric variations of the order of the millimag \citep{Butler:2004}. \cite{Demory:2007} find that the photometric irregularities in the light curve of GJ~436, detected with the EULER Telescope, are due to the presence of a unocculted starspot and they compute a spot filling factor of 1\%, which is compatible with the dispersion in HARPS radial velocity. Thus we set $f_{0} = 0.01$. 

    In Table~\ref{t:depth} we list the obtained results: from the left to the right the planet name, $\Delta R_{\rm p}$/$R_{\rm p}$, and the relative variation in the radius uncertainty, $\Delta$Err, which is the ratio of the computed $\Delta R_{\rm p}$ with respect of the uncertainties quoted in the literature\footnote{See site http://exoplanet.eu/index.php.} (see also Table~\ref{t:cases}). By a simple check, we notice that the largest values of $\Delta R_{\rm p}$/$R_{\rm p}$ and $\Delta$Err occur for the most active star within our sample, i.e. CoRoT-7 with $f_{0} = 0.04$, while the lowest values are found for the coolest and less active stars, i.e. HAT-P-20 and GJ 436 with $f_{0}$ = 0.01 in both cases. For CoRoT-7b, we compute an overestimate of the planetary radius, $\Delta R_{\rm p}$, of about 180 km at optical wavelengths (with a relative variation in the radius uncertainty of about 31\%), while in the K-band it is about 92 km and even smaller in \textit{Spitzer}/IRAC wavebands (at 3.6 $\mu$m, $\Delta R_{\rm p} \simeq 84$~km and at 8.0 $\mu$m $\Delta R_{\rm p} \simeq 75$~km). Such a decrease in $\Delta R_{\rm p}$ as wavelength increases is exhibited by the other examined planetary systems and could be mis-interpreted as evidence of a planetary atmosphere with a wavelength-dependent absorption. Since $\Delta R_{\rm p}$ is proportional to the planetary radius, we obtain the largest $\Delta R_{\rm p}$ for CoRoT-12b ($\Delta R_{\rm p} \simeq 1180$~km at optical wavelengths) with a Jupiter-like planet and the lowest for GJ~436b ($\Delta R_{\rm p} \simeq 130$~km in the optical) with its Super-Earth planet. Nevertheless, all the radius increments calculated in this work fall inside the error bars quoted in the literature, thus no correction to the presently determined planetary parameters is required. We see that for each value of the filling factor, the curves seem to start nearly at the same level (i.e. $\Delta R_{\rm p}$/$R_{\rm p}$ $\simeq$ 0.005 for $f_{0}$ = 0.01 and $\Delta R_{\rm p}$/$R_{\rm p}$ $\simeq$ 0.24 for $f_{0}$ = 0.5), independently of the stellar photospheric temperature. Their trends keep on to be monotonic but their slopes decrease towards cooler photospheric temperatures (cf. GJ~436, in the bottom right panel of Fig.~\ref{f:depth}). This indicates that the overestimate of the planetary radius depends on the wavelength, with the dependence being steeper for solar-type stars and flatter for later spectral types.
    \begin{table*}[!htbp]
\caption{Corrected planetary radii in the optical and Spitzer wavebands for three notable examples: HD 209458b, HD 189733b and GJ 436b.}
\label{t:test}
\centering
\begin{tabular}{cccccc}
\hline
\hline
Planetary & Bands & R$_{corr}$ & R$_{corr}$ & R$_{lit}$ & Ref. \\
Systems & & (R$_{J}$) & (R$_{J}$) & (R$_{J}$) & \\
\hline
\multirow{6}*{\textit{HD~209458b}} & & $f_{0}$ = 0.004 & ($f_{0}$ = 0.03) & & \\
& V & 1.347 & (1.335) & 1.349$\pm$0.022 & (1) \\
& 3.6 $\mu$m & 1.351 & (1.347) & 1.351$\pm$0.006 & (2) \\
& 4.5 $\mu$m & 1.355 & (1.350) & 1.355$\pm$0.008 & (2) \\
& 5.8 $\mu$m & 1.386 & (1.382) & 1.387$\pm$0.007 & (2) \\
& 8.0 $\mu$m & 1.381 & (1.377) & 1.381$\pm$0.005 & (2) \\
\hline
\multirow{6}*{\textit{HD~189733b}} & & ($f_{0}$ = 0.01) & $f_{0}$ = 0.03 & & \\
& V & (1.200) & 1.189 & 1.205$\pm$0.003 & (3) \\
& 3.6 $\mu$m & (1.182) & 1.177 & 1.184$\pm$0.002 & (4) \\
& 4.5 $\mu$m & (1.191) & 1.186 & 1.194$\pm$0.002 & (4) \\
& 5.8 $\mu$m & (1.184) & 1.179 & 1.186$\pm$0.004 & (4) \\
& 8.0 $\mu$m & (1.182) & 1.177 & 1.184$\pm$0.003 & (4) \\
\hline
\multirow{3}*{\textit{GJ~436b}} & & $f_{0}$ = 0.01 & ($f_{0}$ = 0.03) & & \\
& V & 0.400 & (0.396) & 0.402$_{-0.027}^{+0.036}$ & (5) \\
& 8.0 $\mu$m & 0.378 & (0.375) & 0.379$\pm$0.018 & (6) \\
\hline
\end{tabular}
\tablebib{(1)~\citet{Knutson:2007a}; (2) \citet{Beaulieu:2010}; (3) \citet{Pont:2007}; (4) \citet{Desert:2009}; (5) \citet{Bean:2008}; (6) \citet{Deming:2007}.
}
\end{table*}

    Specifically considering the {\it Spitzer} passbands, we apply the method described above to two of the most studied planetary systems: HD~209458 and HD~189733. These systems have been the objects of intense studies in a wide range of wavelengths: from optical with the HST \citep{Brown:2001a,Knutson:2007a,Pont:2007,Sing:2011b} to infrared with the Spitzer satellite \citep{Richardson:2006,Ehrenreich:2007,Tinetti:2007b,Beaulieu:2008,Beaulieu:2010,Desert:2011a}, also as targets for transmission spectroscopy thanks to their brightness.

    In Table~\ref{t:test} we report the results and present the corresponding ones for GJ~436b for comparison. For all the cases, we have considered two different filling factor values of the unocculted spots: one is that argued or reported in the literature, while the other one is given for comparison (indicated in parenthesis in the Table).

    HD~209458 is an almost quiet solar-like G0V star and the calculations have been made considering two configurations: $f_{0} = 0.004$ and $f_{0} = 0.03$. The former $f_{0}$ value is estimated assuming a circular spot with an average radius of $4.5 \times 10^{4}$ km as in the model proposed by \cite{Silva:2003} to fit the distortion of a single transit of HD~209458 observed by HST. HD~189733 is an active early-K star with $f_{0} = 0.03$ \citep{Desert:2011a}. To derive the quantities we have interpolated the BaSeL spectra at temperatures corresponding to the stellar effective temperatures ($T_\mathrm{eff}$ = 6000$~\mathrm{K}$ and $T_\mathrm{eff}$ = 5000$~\mathrm{K}$ for  HD~209458 and HD~189733, respectively) and assuming $\Delta T$ = 1250~$\mathrm{K}$ for both cases. The Table also lists the measured planetary radius and the related uncertainty, as reported in the literature.
    \begin{figure}[!htbp]
     \centering
     \resizebox{\hsize}{!}{\includegraphics[width=8.cm]{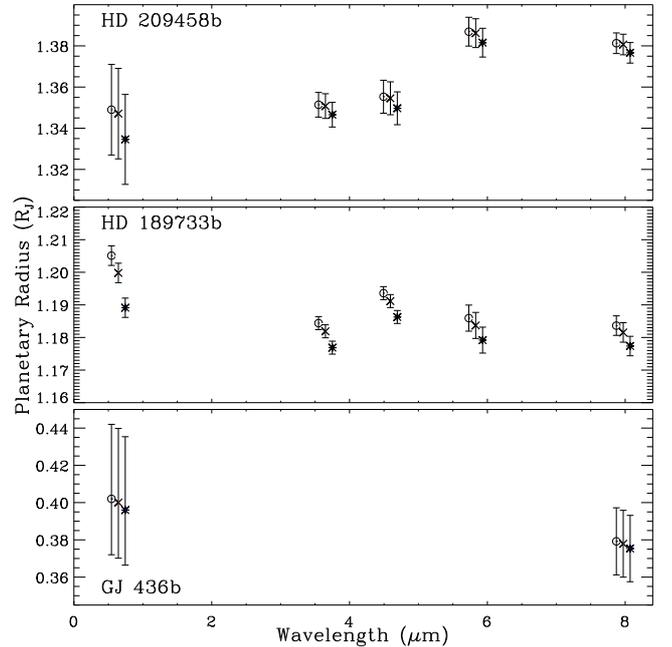}}
     \caption{\small Corrected Planetary Radii vs. Wavelength. \textit{Top panel}, HD 209458b. \textit{Middle panel}, HD~189733b. \textit{Bottom Panel}, GJ~436b. The open $\circ$ circle indicates the planetary radius inferred from the literature, the $\times$ cross the corrected radius with $f_{0} = 0.004$ for HD~209458b and $f_{0} = 0.01$ for HD~189733b and GJ~436b and the $\ast$ asterisk the corrected radius with $f_{0} = 0.03$. The plotted points are shifted horizontally for an easier identification.}
     \label{f:test_radius}
    \end{figure}

    \begin{figure}[!htbp]
     \centering
     \resizebox{\hsize}{!}{\includegraphics[width=9.0cm]{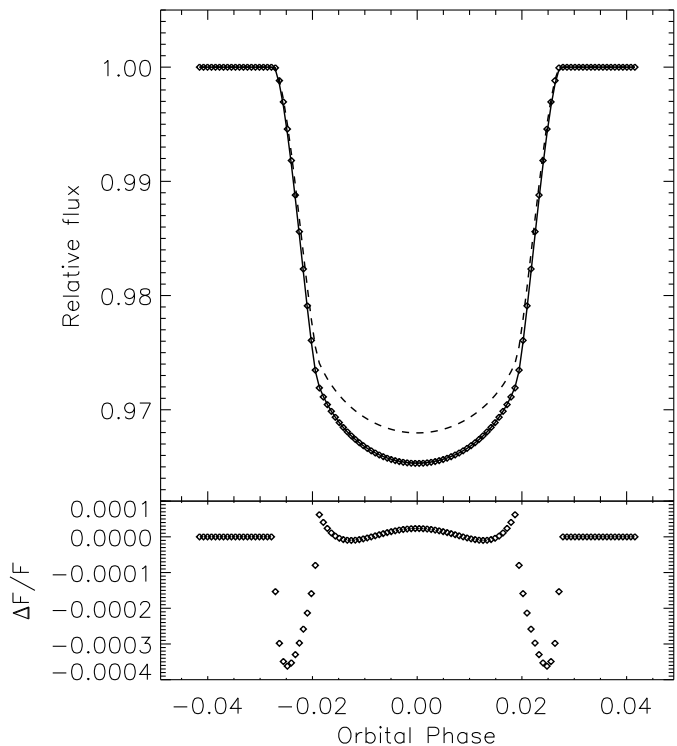}}
     \caption{\small \textit{Top panel}: the synthetic light curve of the transit of CoRoT-2 as computed with the parameters in Table~1 of \citet{Alonso:2008} (dashed line) together with the unperturbed profile obtained from Eq.~(\ref{xx}) (diamonds) and the corresponding best fit (solid line). The flux is measured in units of the out-of-transit flux level in all the cases. \textit{Bottom panel}: the residuals of the best fit to the unperturbed transit profile in relative flux units vs. the orbital phase.}
     \label{mod_transit}
    \end{figure}

    In Fig.~\ref{f:test_radius} we show the  corrected planetary radii  vs. wavelength, along with the uncertainties reported in the literature. The adopted stellar radii are $R_{\rm HD~209458}$ = $1.146$ $R_{\rm \odot}$ \citep{Brown:2001a}, $R_{\rm HD~189733}$ = $0.788$ $R_{\rm \odot}$ \citep{Baines:2008}, $R_{\rm GJ~436}$ = $0.464 R_{\rm \odot}$ \citep{Torres:2007}, respectively. The most active star (i.e. HD~189733) shows significant discrepancies between the planet radius inferred from the literature and the one determined here; this is an expected result since $\Delta R_{\rm p} \propto f$ (see Eq.~\ref{eq:depth2} in Sect.~\ref{radius}) and may have interesting consequences if one wants to characterize the planet atmosphere through transmission spectroscopy. Similarly, in the quiet cases, even though our corrections are within the error bars reported in the literature (of the order of the hundreth or of the thousandth, see Table~\ref{t:test}), they may become significant on the determination of the physical properties of similar exoplanets. The wavelength dependence of the apparent radius of HD~189733b in the optical passbands has been observed by \citet{Sing:2011b}. They used an approach similar to that of Sect. \ref{radius} and concluded that the variation of $R_{\rm p}/R_{\star}$ vs. wavelength $\lambda$ cannot be reproduced by an equation analogue to Eq. (\ref{eq:rvsl}), thus providing evidence of a wavelength dependence of the planetary atmospheric absorption.

  \subsection{Effects of stellar activity on orbital parameters}
   \label{star_parameter}
   To illustrate the impact of starspots on the  determination of all the transit parameters, we consider the case of CoRoT-2 because it has been already studied by \citet{Czesla:2009} and its spot activity has been investigated in detail. Since we are interested in the systematic errors produced by starspots, we simulate a noiseless transit profile adopting the time sampling and the parameters as derived by \citet{Alonso:2008}, and apply Eq.~(\ref{xx}) with spot filling factors estimated from  \citet{Lanza:2009} for the out-of-transit light curve and \citet{Silva-Valio:2010b} for the transit light curve, respectively. To derive the transit profile to be fitted with their model, \citet{Alonso:2008} averaged all the transits observed by CoRoT. Therefore, we assume that the spot perturbation corresponds to the average values of the filling factors $f_{0}$ and $f_{\rm i}$ as derived from the modelling of the out-of-transit and the transit light curves, respectively. By fitting the light curve of CoRoT-2 outside transits, \citet{Lanza:2009} derived an average spot filling factor $f_{0} = 0.07$ adopting a spot contrast in the CoRoT white passband of $A_{\rm LANZA} = 0.665$. On the other hand, \citet{Silva-Valio:2010b} found an average filling factor of $f_{\rm i} = 0.15$ for a spot contrast of $A_{\rm SILVA} = 0.55$. Multiplying it by the ratio $A_{\rm LANZA}/A_{\rm SILVA}$ we convert it into the scale of \citet{Lanza:2009}, giving $f_{\rm i} = 0.18$. With those values of the filling factors and $A_{\lambda} = 0.665$, we compute the unperturbed transit profile $D_{0}$ from Eq.~(\ref{xx}) and then fit it with the Levenberg-Marquardt algorithm as implemented in the IDL procedure {\sc lmfit}. The unperturbed transit profile and its best  fit are plotted in Fig.~\ref{mod_transit} together with the initial profile assumed to be perturbed by the effects of the spots. The best fit has a reduced $\chi^2=1.29$ computed by assuming a relative standard deviation of the simulated data of $1.09 \times 10^{-4}$ as in the case of the CoRoT data analysed by \citet{Alonso:2008}. The best fit parameters with their  standard deviations as derived from the covariance matrix with the above assumed data standard deviation \citep[see ][ \S15.4]{Press:1992} are: $a/R_{\star} = 6.76 \pm 0.03$, $R_{\rm p}/R_{\star} = 0.1726 \pm 0.00006$, $i = 88\fdg 44 \pm 0\fdg18$, $u_{1}=0.40 \pm 0.04$, and $u_{2} = 0.097 \pm 0.04$. 

   We confirm that the radius variation is close to that estimated with the simple approach of Sect.~\ref{radius} because the depth of the profile at mid transit gives the principal constraint on the relative radius. However, contrary to the results of \citet{Czesla:2009}, we find that the relative semimajor axis $a/R_{\star}$ and the inclination $i$ are also affected, with variations that exceed three standard deviations for both parameters \citep[cf. Table~1 of ][]{Alonso:2008}. Note that the transit duration $t_{\rm T}$  is the same in our model for both the unperturbed and the perturbed transit profiles. Therefore, the variations of $a/R_{\star}$ and $i$ combine with each other as to maintain $t_{\rm T}$ practically constant (see Sect.~\ref{param_var}). The variation of $a/R_{\star}$ has an impact on the measurement of the density of the star \citep[cf. ][]{Seager:2003,Sozzetti:2007} that, together with its effective temperature, is used to derive its position on the H-R diagram and hence its age from the isochrones. We shall not consider this application any further because it is beyond the scope of the present work, but conclude that starspot effects cannot be neglected in the derivation of the stellar density from the transit fitting in the case of active planetary hosts.

   The limb-darkening parameters are not significantly affected. The small systematic deviations in the ingress and egress of the profile do not lead to any significant adjustment of the limb darkening (cf. upper and lower panels of Fig.~\ref{mod_transit}).

   Other stellar parameters can be affected by starspots, notably the effective temperature $T_{\rm eff}$. The problem has been discussed in detail in the case of CoRoT-2 by \citet{Guillot:2011} who found that the effect of  spots is that of increasing the uncertainty in the determination of $T_{\rm eff}$ because of their changing covering factor along the activity cycle of the star.  We shall not consider further these aspects because here we focus on the parameters directly related to the modelling of the transit light curve.

 \section{Conclusions}
  The observation of moderately active stars simultaneously in the optical and in the near infrared passbands allows us to discriminate  between planetary transits and activity-induced variations, unless starspots are  $2000-2500$ $\mathrm{K}$ cooler than the unperturbed photosphere, which is generally not observed in main-sequence stars. This approach can be applied to individual light curve dips observed with a signal-to-noise ratio of $\approx$ $10 - 15$\% of the flux variations in  two well-separated passbands, e.g., U and K, and does not require a characterization of the stellar microvariability timescales which implies observations extended for at least $2-3$ stellar rotations. 

  Moreover, infrared observations are important for the  derivation of the planetary radius and accordingly of the planetary density. The correction of the effects due to starspots is necessary to avoid systematic errors in trasmission spectroscopy, especially when comparing the planet radius at optical and infrared passbands. A correct determination of the planetary radius  allows us a better study of the planetary atmosphere and its  compositions in the framework of the transmission and reflection spectroscopy. We provide simple formulae, tabulations, and graphics to help estimating the starspot perturbation in the case of typical transiting systems. Moreover, we find that starspots may affect the relative semimajor axis $a/R_{\star}$ and  the inclination $i$ as derived  by transit fitting. This may have a non-negligible impact on the determination of the stellar density, especially when optical light curves are used. Near infrared light curves are less affected, but the effect may still be significant for very active stars.

  A simultaneous broad-band photometric observation, from visible to infrared wavelengths, of a significant sample of stars, on one hand might reveal itself as a relevant tool to identifying and characterizing their stellar activity, allowing us to disentagle its contribution from  planetary transit signals, on the other hand would help us to derive relevant and more accurate information on the primary parameters of the planets (as already suggested by \citealt{Jha:2000}) and the physical properties of their atmospheres, such as better constraints on temperatures and compositions.

\begin{acknowledgements} The authors gratefully acknowledge an anonymous referee and the A\&A editor Dr. T. Guillot for their valuable comments on a previous version of this manuscript that greatly helped to improve their work. Support for this research has been provided by the contract PRIN-INAF (P.I.: Lanza). We acknowledge contribution from ASI (agreement I/044/10/0).
\end{acknowledgements}

\bibliographystyle{aa}
\bibliography{biblio}

\end{document}